\documentclass[aip,jap,reprint,amsmath,amssymb,superscriptaddress,floatfix,citeautoscript]{revtex4-2}
\usepackage{graphicx}
\usepackage{color}
\usepackage{bm}
\usepackage{braket}
\usepackage[colorlinks,linkcolor=blue,citecolor=blue,urlcolor=blue]{hyperref}

\newcommand{\gao}{Ga$_2$O$_3$}
\newcommand{\bgao}{$\beta$-Ga$_2$O$_3$}

\begin{document}

\title{Characterization of Chromium Impurities in {\bgao}}

\author{Mark E. Turiansky}
\email{mturiansky@ucsb.edu}
\altaffiliation[Present Address: ]{US Naval Research Laboratory, 4555 Overlook Avenue SW, Washington, DC 20375, USA}
\affiliation{Materials Department, University of California, Santa Barbara, CA 93106-5050, U.S.A.}

\author{Sai Mu}
\email{MUS@mailbox.sc.edu}
\affiliation{Materials Department, University of California, Santa Barbara, CA 93106-5050, U.S.A.}
\affiliation{Department of Physics and Astronomy, University of South Carolina, Columbia, South Carolina, 29208, U.S.A.}

\author{Lukas Razinkovas}
\affiliation{Center for Physical Sciences and Technology (FTMC), Vilnius LT--10257, Lithuania}
\affiliation{Department of Physics/Centre for Materials Science and Nanotechnology, University of Oslo, P.O. Box 1048, Blindern, Oslo N-0316, Norway}

\author{Kamyar Parto}
\affiliation{Department of Electrical and Computer Engineering, University of California, Santa Barbara, CA 93106-5080, U.S.A.}

\author{Sahil D. Patel}
\affiliation{Department of Electrical and Computer Engineering, University of California, Santa Barbara, CA 93106-5080, U.S.A.}

\author{Sean Doan}
\affiliation{Department of Physics, University of California, Santa Barbara, CA 93106-9530, U.S.A.}

\author{Ganesh Pokharel}
\affiliation{Materials Department, University of California, Santa Barbara, CA 93106-5050, U.S.A.}
\affiliation{Perry College of Mathematics, Computing, and Sciences, University of West Georgia, Carrollton, GA 30118, U.S.A.}

\author{Steven J. Gomez Alvarado}
\affiliation{Materials Department, University of California, Santa Barbara, CA 93106-5050, U.S.A.}

\author{Stephen D. Wilson}
\affiliation{Materials Department, University of California, Santa Barbara, CA 93106-5050, U.S.A.}

\author{Galan Moody}
\affiliation{Department of Electrical and Computer Engineering, University of California, Santa Barbara, CA 93106-5080, U.S.A.}

\author{Chris G. \surname{Van de Walle}}
\email{vandewalle@mrl.ucsb.edu}
\affiliation{Materials Department, University of California, Santa Barbara, CA 93106-5050, U.S.A.}

\date{\today}

\begin{abstract}
    Chromium is a common transition-metal impurity that is easily incorporated during crystal growth.
    It is perhaps best known for giving rise to the 694.3~nm (1.786~eV) emission in Cr-doped Al$_2$O$_3$, exploited in ruby lasers.
    Chromium has also been found in monoclinic gallium oxide, a wide-bandgap semiconductor being pursued for power electronics.
    In this work, we thoroughly characterize the behavior of Cr in Ga$_2$O$_3$ through theoretical and experimental techniques.
    {\bgao} samples are grown with the floating zone method and show evidence of a sharp photoluminescence signal, reminiscent of ruby.
    We calculate the energetics of formation of Cr from first principles, demonstrating that Cr preferentially incorporates as a neutral impurity on the octahedral site.
    Cr possesses a quartet ground-state spin and has an internal transition with a zero-phonon line near 1.8~eV.
    By comparing the calculated and experimentally measured luminescence lineshape function, we elucidate the role of coupling to phonons and uncover features beyond the Franck-Condon approximation.
    The combination of strong emission with a small Huang-Rhys factor of 0.05 and a technologically relevant host material render Cr in Ga$_2$O$_3$ attractive as a quantum defect.
\end{abstract}

\maketitle

\section{Introduction}
Monoclinic {\gao} ({\bgao}) has garnered attention for its ability to be controllably $n$-type-doped despite its ultra-wide ($\sim$4.9~eV) bandgap~\cite{tsao_ultrawide-bandgap_2018,mohamed_surface_2011}.
Its high breakdown field and availability of high-quality low-cost substrates make it an excellent material for high-power electronics and ultraviolet optoelectronics~\cite{tsao_ultrawide-bandgap_2018,alema_solar_2019,oshima_vertical_2008,higashiwaki_gallium_2012}.
These properties also render {\bgao} a suitable host for quantum defects~\cite{dreyer_first-principles_2018,turiansky_rational_2024}. 
Realizing any of these applications in {\bgao} demands control over the native point defects and impurities present in the crystal and a deeper understanding of their electrical and optical properties relevant for spin and photonic qubits~\cite{azzam_prospects_2021}.

Numerous defects and impurities have been uncovered in {\bgao}~\cite{varley_oxygen_2010,mu_role_2022,frodason_self-trapped_2020,tadjer_editors_2019,mccluskey_point_2020}.
Chromium is a common impurity, well known for its role in ruby---corundum-Al$_2$O$_3$ doped with Cr---that enabled the first solid-state lasers~\cite{maiman_stimulated_1960}.
Cr in Al$_2$O$_3$ is also being explored as a potential spin qubit~\cite{sewani_spin_2020}.
The corundum and monoclinic crystal structures have similarities, such as the presence of an octahedrally coordinated cation site;
however, the lower-symmetry monoclinic crystal also includes a tetrahedrally coordinated cation site.
We thus expect the incorporation and behavior of Cr in {\bgao} to have complexities beyond those observed in ruby.

Signatures of Cr incorporation in {\bgao} have been reported.
Electron-paramagnetic resonance (EPR) experiments found spin-active centers associated with Cr in the 3+ oxidation state~\cite{tippins_optical_1965,yeom_electron_2003,stehr_magneto-optical_2021}.
Deep-level transient spectroscopy (DLTS) and thermoluminescence (TL)~\cite{luchechko_effect_2020,esteves_probing_2023} have found deep levels suggested to be due to Cr.
Photoluminescence spectra reminiscent of ruby have also suggested to be due to Cr~\cite{stehr_magneto-optical_2021,tippins_optical_1965,sun_origin_2020,barmore_photoluminescence_2023},
and a long lifetime ($\approx$ms) was observed~\cite{tippins_optical_1965}.
Given the many uncertainties that are still associated with these attributions to Cr in {\bgao}, a thorough characterization of its incorporation as well as its electronic and optical properties is warranted.

Here we combine advanced first-principles calculations with photoluminescence measurements on {\bgao} grown by the floating zone method to assess the properties of Cr in {\bgao}.
Utilizing density functional theory (DFT) with a hybrid functional, we find that Cr preferentially incorporates on the octahedral cation site, giving rise to several levels in the bandgap.
We focus on Cr in a 3+ oxidation state, which corresponds to the neutral charge state of the impurity and is stable for most of the relevant range of Fermi levels. 
We study the excited-state structure and find an internal transition between a $^2E$ excited state and $^4A_2$ ground state near 1.8~eV.
We simulate the photoluminescence spectrum from first principles and compare with explicit measurements on {\bgao} samples grown by the floating zone method.
The agreement provides compelling evidence that the observed spectrum, with a zero-phonon line (ZPL) at 1.79 eV, is indeed due to substitutional Cr on an octahedral site.
We theoretically and experimentally verify a small Huang-Rhys factor of $\approx$0.05, which in combination with a wide bandgap, millisecond-long recombination lifetime, and sharp ZPL make this center a compelling candidate for spin and photonic qubits.

The paper is organized as follows.
In Sec.~\ref{sec:growth}, we discuss the details of the floating zone growth of monoclinic {\bgao}.
In Sec.~\ref{sec:pl_measure}, we cover the measurements of the photoluminescence spectrum of the {\bgao} samples in the visible spectrum, including polarization dependence.
In Sec.~\ref{sec:dft}, we outline the computational methodology and results employed to study Cr impurities in {\bgao}.
We discuss the implications of all these results in Sec.~\ref{sec:discuss} and conclude in Sec.~\ref{sec:concl}.

\section{Growth of {\bgao}}
\label{sec:growth}
A single crystal sample of {\bgao} was grown via the floating zone method, using a locally designed laser floating zone furnace ``LOKII'' in the Wilson lab at UC Santa Barbara~\cite{gomez_alvarado_advances_2024}.
In preparation for the crystal growth, polycrystalline powder of {\gao} (99.99 $\%$, Alfa Aesar) was dried overnight at 1200\,$^\circ$C and then packed in a rubber balloon with a diameter of 4~mm.
The packed powder in the balloon was pressed into a 90~mm long rod using a cold isostatic press. The rod was then broken into two pieces, measuring 65~mm and 25~mm in length.
These two pieces served as the feed rod and seed rod for the growth process.
Subsequently, the rods were sintered at 1200\,$^\circ$C for 10 hours in air to increase their density and hardness.

\begin{figure}
    \centering
    \includegraphics[width=\columnwidth,height=0.5\textheight,keepaspectratio]{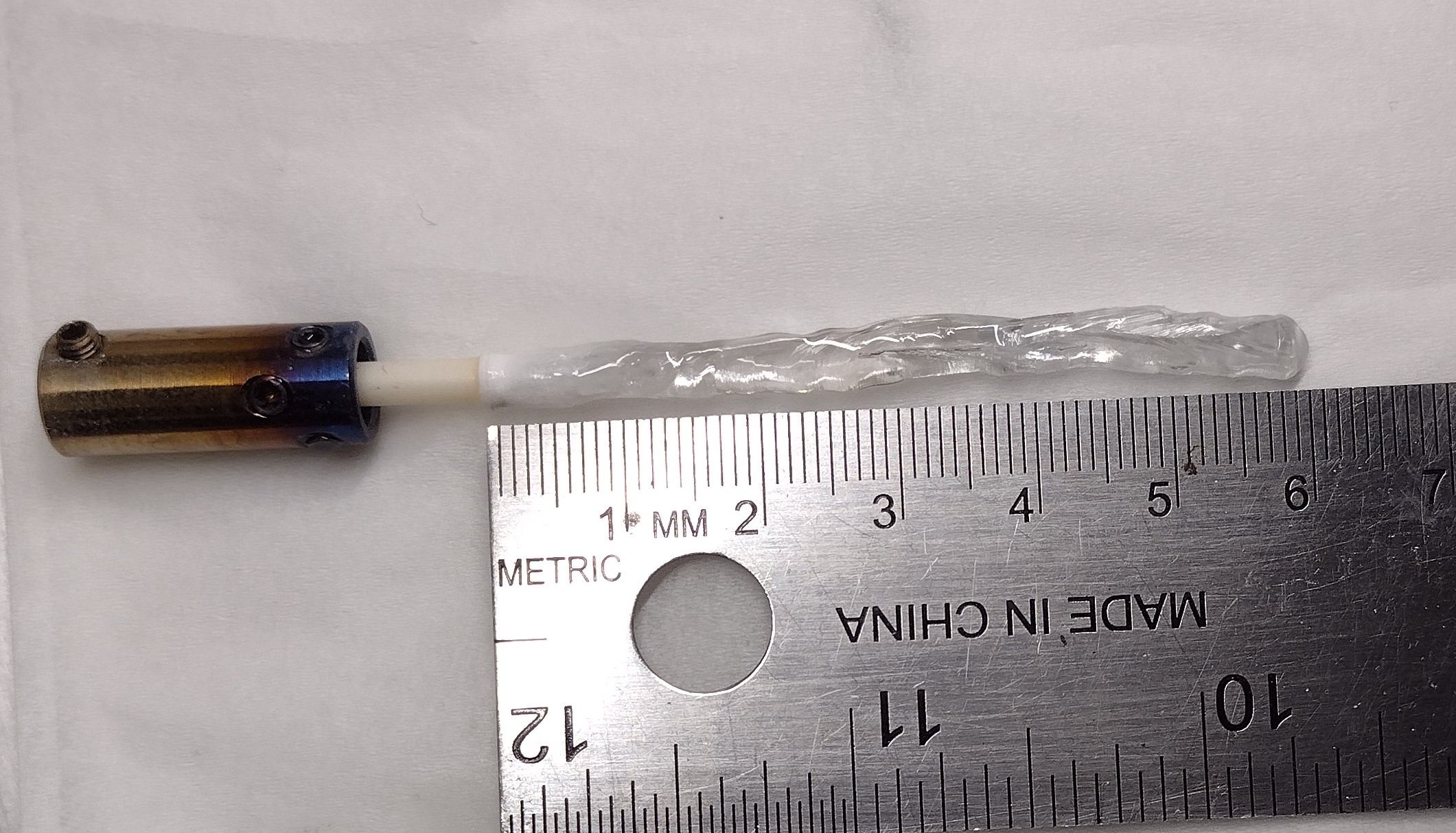}
    \caption{\label{fig:growth}
        Single crystal of {\bgao} grown from high-pressure laser floating zone furnace.
    }
\end{figure}

An initial melting test with the seed rod of {\bgao} showed that an incident laser power of 100~W per laser (600~W total) was not sufficient to melt the seed rod.
Therefore, an alumina rod was used as a seed rod, which melted at around 1900\,$^\circ$C using a laser power of 36~W per laser (216~W total).
The growth of {\bgao} was conducted using a pressure of 1000~psi of an Ar:O$_2$ = 80:20 mix to prevent any volatility.
Growth conditions were kept constant throughout the process, including a translation rate of 15~mm/hr for both the seed and feed rods, a rotation rate of 10 rpm, and a total incident laser power of~168 W.
The length of the single crystal of {\bgao} is 60~mm as shown in Fig.~\ref{fig:growth}.
The phase purity of the crystal was confirmed by measuring a micron-sized broken piece of the rod using a Kappa Apex II single-crystal diffractometer.
However, due to the experimental limitations, we cannot rule out the presence of alternative phases making up less than 1\% of the crystal.
Additionally, single-crystal Laue diffraction was collected using the horizontal Laue System AK261, manufactured by Photonic Science, to determine the crystal orientation.

Any incorporation of Cr in the sample was unintentional.
We believe there are two possible sources of Cr contamination during the growth.
X-ray diffraction shows evidence for the presence of CrO$_2$ in the polycrystalline Ga$_2$O$_3$ source material.
In addition, the alumina seed rod was relatively low purity and may also be a source of contamination.

\section{Photoluminescence Spectroscopy}
\label{sec:pl_measure}
To characterize the optical properties of the grown {\bgao} sample at wavelengths in the visible spectrum, photoluminescence spectroscopy was performed at 4.2~K using a 532~nm laser diode as an excitation source.
A linear polarizer and half-wave plate in the excitation path were utilized to set the excitation of the input laser.
Reflected light is rerouted through the collection path where another set of half-wave plate and linear polarizer filter the polarization of the collected light.
A low-pass filter at 550~nm is used to block the residual laser light.
Light is finally routed to a Princeton Instruments spectrometer.
Dominant ZPL peaks were measured using a 300~groove/mm grating, while fine features of the phonon sideband were measured using an 1800~groove/mm grating.

\begin{figure}
    \centering
    \includegraphics[width=\columnwidth,height=0.5\textheight,keepaspectratio]{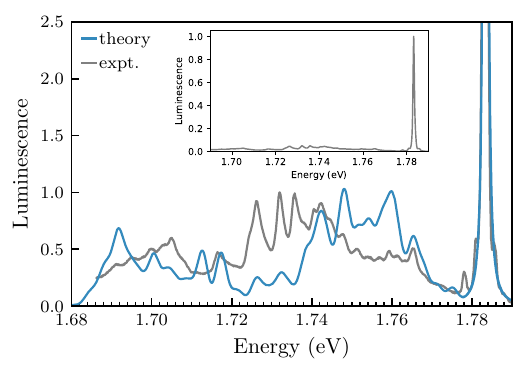}
    \caption{\label{fig:lineshape}
        Experimental (gray) and theoretical (blue) luminescence lineshapes.
        The theoretical lineshape is normalized and expressed in units of meV$^{-1}$;
        the ZPL is shifted to match experiment.
        The experimental lineshape is scaled to match the theoretical lineshape.
        The inset shows a zoomed out view of the experimental lineshape normalized to the peak intensity, which emphasizes the emission into the ZPL.
    }
\end{figure}

The measured photoluminescence lineshape function is shown in Fig.~\ref{fig:lineshape}.
The spectrum results from ensemble emission;
isolating a single emitter was not possible, given the overall low intensity consistent with the long lifetime of emission.
A sharp line at 1.79~eV associated with the zero-phonon line is found, as well as several smaller peaks at lower energy.
The observed photoluminescence peaks did not show any significant power dependence.
Given the similarity of the emission to that of Cr in Al$_2$O$_3$, as well as the similarity between the two crystal structures, we hypothesize that the emission is due to unintentional Cr incorporation during the growth.
We will confirm this hypothesis in the Sec.~\ref{sec:dft} by comparing with a spectrum calculated with DFT.

\begin{figure*}
    \centering
    \includegraphics[width=\textwidth,height=0.5\textheight,keepaspectratio]{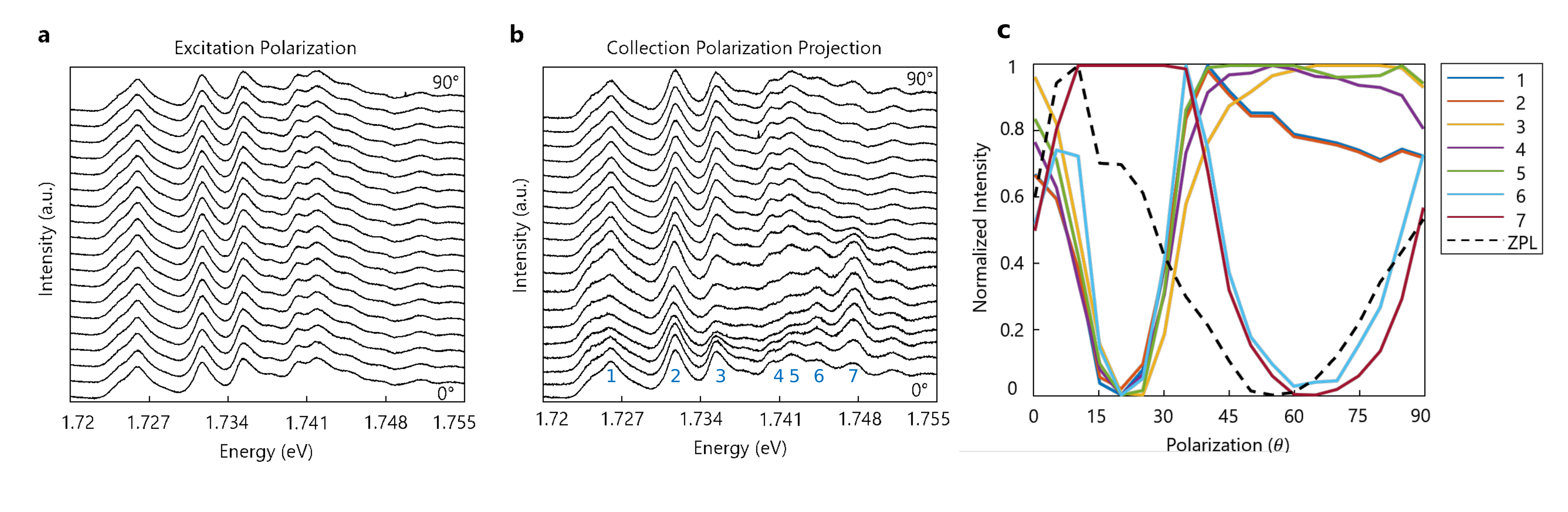}
    \caption{\label{fig:polarization}
        The phonon sideband as a function of the (a) excitation and (b) emission polarization angle.
        (c) The normalized intensity of several peaks in the phonon sideband [labeled in panel (b)] along with the ZPL as a function of the emission polarization angle.
    }
\end{figure*}

To further understand the properties of the luminescent center, we measure the polarization dependence of the photoluminescence spectra as shown in Fig.~\ref{fig:polarization}.
Excitation polarization spectroscopy was performed by rotating the half-wave plate at the excitation path and removing the polarizers on the collection path.
Similarly, emission polarization was measured by rotating the half-wave plate in the collection path.
We find that each peak of the photoluminescence spectrum exhibits the same dependence on the excitation polarization angle [Fig.~\ref{fig:polarization}(a)].
In contrast, the spectrum changes significantly as a function of the polarization of the emitted light [Fig.~\ref{fig:polarization}(b)].
This observation is unexpected within the usual Franck-Condon approximation~\cite{lax_franckcondon_1952} of photoluminescence, in which all peaks in the phonon sideband should behave the same as the ZPL.

To be more quantitative, we plot the normalized intensity of several peaks in the phonon sideband along with the ZPL as a function of the emission polarization angle in Fig.~\ref{fig:polarization}(c).
We see that the peaks labeled as 1--3, which correspond to the three peaks discussed in the main text, exhibit an opposite dependence from that of the ZPL:
when the ZPL intensity is maximal, these peaks are minimal, and vice versa.

\section{Density Functional Theory}
\label{sec:dft}

\subsection{Energetics of Cr Incorporation}
We perform DFT calculations using the projector augmented-wave method (PAW)~\cite{blochl_projector_1994} implemented in the Vienna \textit{Ab-initio} Simulation Package (VASP)~\cite{kresse_efficiency_1996,kresse_efficient_1996}.
For the exchange-correlation potential, we utilize the hybrid functional of Heyd, Scuseria, and Ernzerhof~\cite{heyd_hybrid_2003,heyd_erratum:_2006} to ensure accurate predictions.
The screening parameter is set to the default value, and we tune the mixing parameter to 32\%, resulting in a bandgap of 4.83~eV in good agreement with the experimental value of 4.9~eV~\cite{mohamed_surface_2011}.
The wavefunctions were expanded in a plane-wave basis with an energy cutoff of 400~eV.
Ga $4s^2p^1$, O $2s^2p^4$, and Cr $4s^13d^5$ were included as valence electrons.
We performed select tests to ensure that the Ga $d$ states could be safely included in the core.
A 120-atom supercell was constructed from a $1\times3\times2$ multiple of the conventional monoclinic cell~\cite{peelaers_brillouin_2015} for the point-defect calculations.
For each point defect and charge state, we perform multiple calculations with different initial local distortions around the defect in the initial state, and optimize the atomic positions until the Hellmann-Feynman forces are lower than 5 meV/\AA\@.
Brillouin-zone integration was carried out using a $\Gamma$-centered 4$\times$4$\times$4 k-point mesh for the primitive cell and 2$\times$2$\times$2 for the supercell.
Spin polarization is explicitly included.

We calculate the formation energy of Cr substituting on a Ga site~\cite{freysoldt_first-principles_2014}:
\begin{equation}
    \begin{split}
        E^f({\rm Cr}^q_{\rm Ga}) & = E_{\rm tot}({\rm Cr}^q_{\rm Ga})-E_{\rm tot}({\rm Ga}_2{\rm O}_3) - \mu_{\rm Cr} \\
        & + \mu_{\rm Ga} + q(E_{\rm F}+E_{\rm VBM}) + \Delta^q \;,
    \end{split}
\end{equation}
where $E_{\rm tot}({\rm Cr}^q_{\rm Ga})$ is the total energy of a supercell containing Cr$_{\rm Ga}$ in charge state $q$, $E_{\rm tot}({\rm Ga}_2{\rm O}_3)$ is the total energy of a bulk supercell, and $E_{\rm F}$ is the Fermi energy referenced to the valence-band maximum (VBM).
$\Delta^q$ is a finite-size correction term for charged defects~\cite{freysoldt_fully_2009,freysoldt_electrostatic_2011}.

$\mu_{{\rm Cr}/{\rm Ga}}$ are the chemical potentials referenced to the appropriate elemental phase (Cr metal or Ga metal).
The chemical potentials describe the growth and processing conditions: we will assume Ga-rich growth conditions and incorporation of Cr at the solubility limit.
Cr$_2$O$_3$ (chromia) is a potential secondary phase that may form when Cr is present during growth and sets the solubility limit for Cr.
Cr$_2$O$_3$ is most stable in the rhombohedral phase and is calculated in the ground-state spin configuration, which is G-type antiferromagnetic.
We note that when Cr$_2$O$_3$ is used as a limiting phase, the formation energy of substitutional Cr is independent of the growth conditions (i.e. Ga-rich vs. O-rich).

\begin{figure}
    \centering
    \includegraphics[width=\columnwidth,height=0.5\textheight,keepaspectratio]{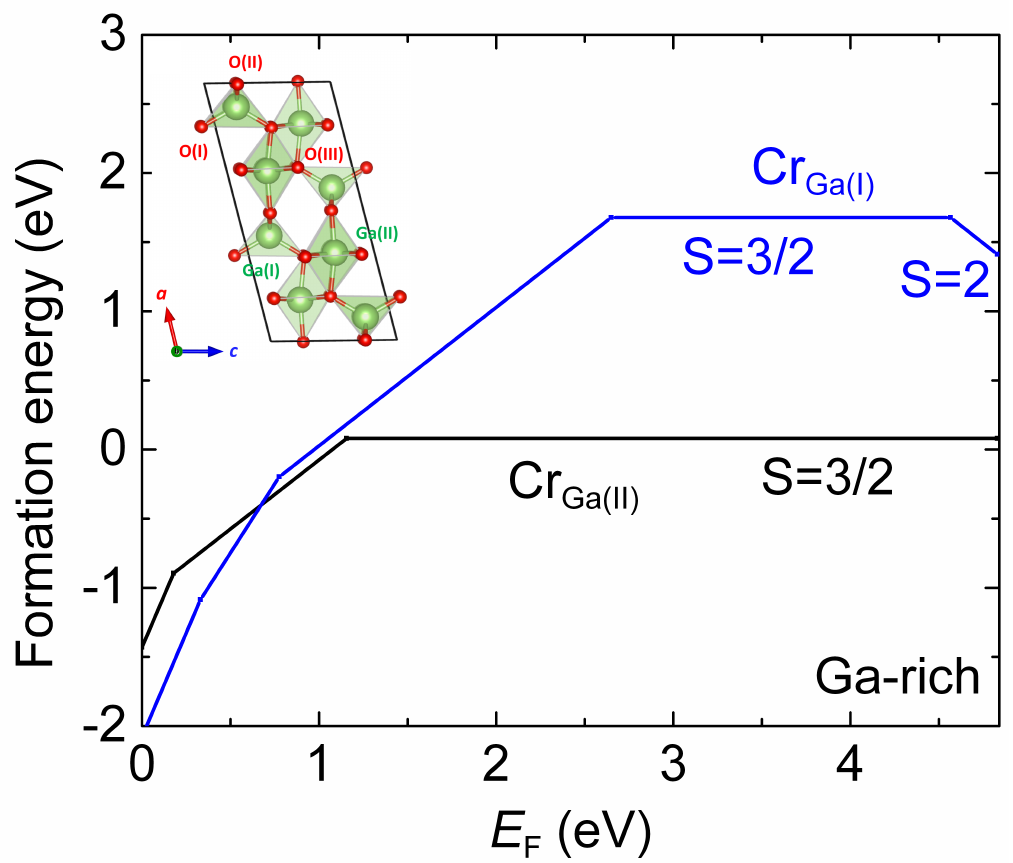}
    \caption{\label{fig:form_en}
        Formation energy versus Fermi level for substitutional Cr$_{\rm Ga}$ in {\bgao} under Ga-rich conditions.
        The total ground-state spin $S$ is denoted for each charge state.
        Inset: primitive cell of the {\bgao} crystal structure.
        O atoms are shown in red and Ga in green.
        Ga and O atoms in different coordinations are labeled with a Roman numeral.
    }
\end{figure}

The calculated formation energies are shown in Fig.~\ref{fig:form_en}.
The {\bgao} crystal structure (inset of Fig.~\ref{fig:form_en}) contains two types of cation sites: the tetrahedral site (denoted as I) and the octahedral site (denoted as II).
O(I) are three-fold coordinated (on a shared corner of two edge-sharing GaO$_4$ octahedra and one GaO$_6$ tetrahedron), and O(II) are also three-fold coordinated (on the shared corner of one GaO$_6$ octahedron and two GaO$_4$ tetrahedra);
O(III), on the other hand, are four-fold coordinated.
We find that Cr prefers incorporating on the octahedral site:
Cr$^0_{\rm Ga(II)}$ is lower in energy than Cr$^0_{\rm Ga(I)}$ by 1.60~eV.
The formation energy of Cr$^0_{\rm Ga(II)}$ is as low as 0.07 eV for Ga-rich conditions, indicating that Cr can be easily incorporated.
This is consistent with the fact that Cr is octahedrally coordinated with oxygen in Cr$_2$O$_3$.
The site preference of Cr also agrees with the findings of experimental observations~\cite{geller_crystal_1960,yeom_electron_2003,stehr_magneto-optical_2021}.
We find a (+/0) transition level for Cr$_{\rm Ga(II)}$ at 1.15 eV above the VBM\@.
Given that acceptors in {\bgao} are all deep, with ionization energies of at least 1.3~eV~\cite{lyons_survey_2018}, we conclude that  Cr$_{\rm Ga(II)}$ will exclusively occur in the neutral charge state, corresponding to the 3+ oxidation state and consistent with EPR measurements~\cite{yeom_electron_2003,stehr_magneto-optical_2021}.

Incorporation of Cr on O sites is unlikely because of the large size mismatch.
We investigated Cr in interstitial configurations but found its formation energy to be much higher than that of Cr$_{\rm Ga(II)}^0$ when the Fermi level is within $\sim$1~eV of the conduction-band minimum (CBM).

For Cr$_{\rm Ga(I)}$ we find a (0/$-$) transition level at 0.27~eV below the CBM, indicating that
the negative charge state (with ground-state spin $S = 2$) is stable under $n$-type conditions.
The stability of a negative charge state for Cr$_{\rm Ga(I)}$ prompted us to explicitly look for it in the case of Cr$_{\rm Ga(II)}$:
we were able to converge a locally stable Cr$^-_{\rm Ga(II)}$ configuration, but the (0/$-$) transition level was found to be 0.70~eV above the CBM, indicating it can never be thermodynamically stabilized.

Focusing on Cr$_{\rm Ga(II)}^0$, there is no notable local distortion in the atomic geometry.
The relaxed Cr-O bond lengths are 1.96 \AA\ for the two Cr-O(I) bonds, 1.94 \AA\ for the two Cr-O(II) bonds, and 1.99 \AA\ and 2.03 \AA\ for Cr-O(III), which are similar to Ga(II)-O bonds in pristine {\gao}.
These results indicate that the octahedral environment is maintained for Cr$_{\rm Ga(II)}^0$;
however, due to the relatively low symmetry of the {\bgao} crystal, we find a local $C_{1h}$ symmetry for Cr$_{\rm Ga(II)}^0$.

\subsection{Ground-State Properties}

\begin{figure}[th]
    \centering
    \includegraphics[width=\columnwidth,height=0.5\textheight,keepaspectratio]{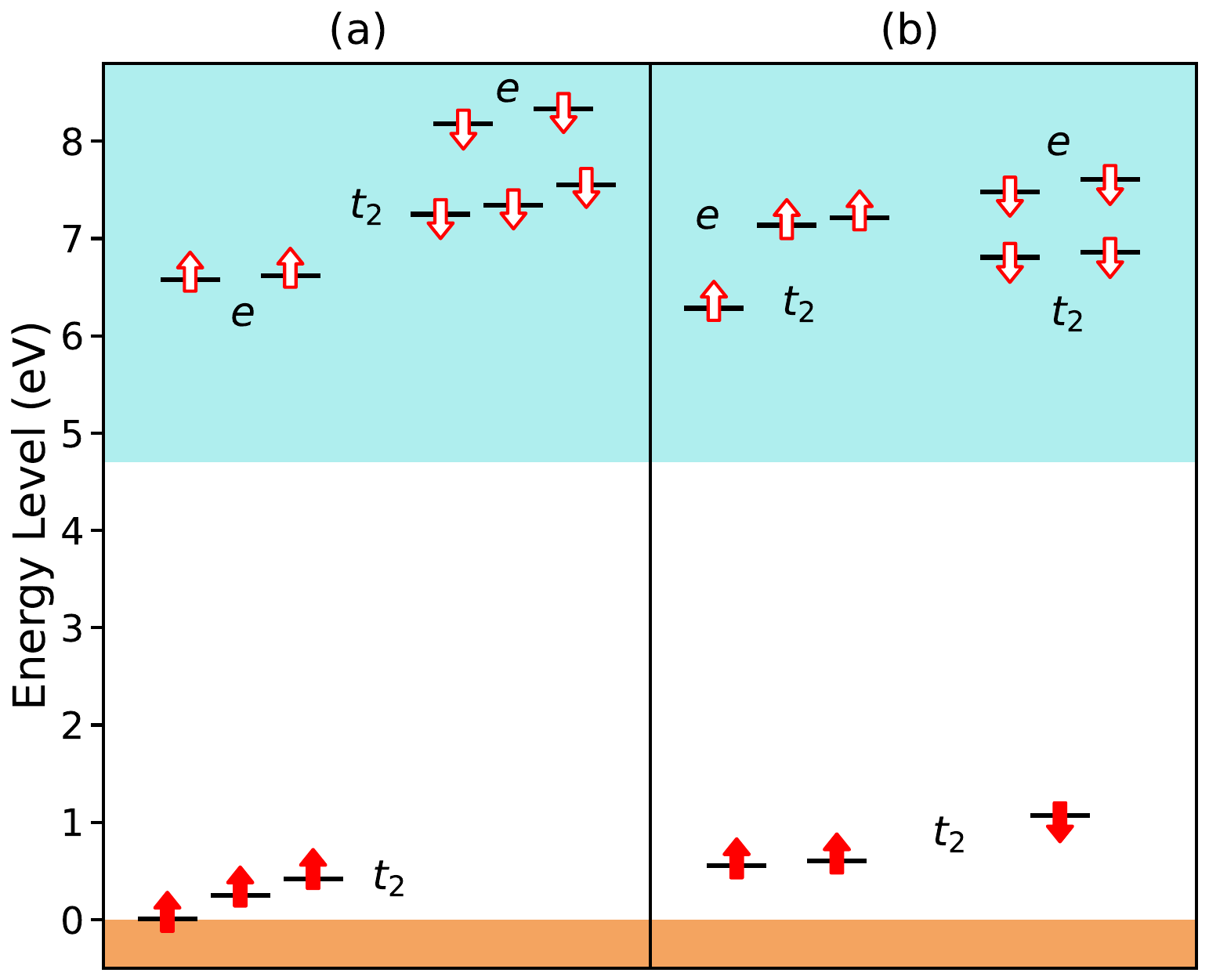}
    \caption{\label{fig:ks}
        HSE Kohn-Sham eigenvalue spectrum for Cr$^0_{\rm Ga(II)}$ in {\bgao} in: (a) the $^4A_2$ ground state; (b) the $^2E$ excited state with a single spin flip.
        The orange (blue) shaded area represents the valence (conduction) band.
        Spin-up and spin-down channels are depicted; occupied (empty) states are indicated by filled (open) red arrows.
        The Fermi level is referenced to the VBM.
    }
\end{figure}

The Kohn-Sham states of Cr$_{\rm Ga(II)}^0$ are shown in Fig.~\ref{fig:ks}(a).
For the neutral charge state, three electrons occupy the $d$ manifold of Cr, giving a $d^3$ configuration.
In a perfect octahedral environment (point group $O_h$), the five-fold degenerate $d$ manifold of Cr splits into a three-fold degenerate $t_2$ and a two-fold degenerate $e$ manifold, with the $t_2$ states lower in energy than the $e$ states.
(As all states have gerade symmetry here, we drop the label.)
Subduction from $O_h$ to $C_{1h}$ results in a lifting of the degeneracy in the $t_2$ and $e$ states.
However, as indicated by the relatively low spread in bond lengths and Kohn-Sham states [Fig.~\ref{fig:ks}(a)], this symmetry breaking is relatively minor, and states with $t_2$ and $e$ character can still be readily distinguished.
In the following analysis, we will therefore utilize the $O_h$ irreducible representation to label the electronic states.
We find a $^4A_2$ ground-state structure for Cr$_{\rm Ga(II)}^0$, consistent with the Tanabe-Sugano diagram for $d^3$ that suggests a quartet ground state independent of crystal-field strength~\cite{sugano_multiplets_1970}.

Even though Cr$_{\rm Ga(I)}^0$ is significantly higher in energy than Cr$_{\rm Ga(II)}^0$, we cannot exclude that it could be incorporated in the crystal under non-equilibrium conditions.
We therefore also examine its electronic structure.
In a tetrahedral crystal field, the $e$ states are lower in energy than the $t_2$ states.
Combined with particle-hole symmetry, a $d^3$ system with tetrahedral symmetry is thus equivalent to a $d^7$ system with octahedral symmetry~\cite{sugano_multiplets_1970}.
We thus expect that Cr$_{\rm Ga(I)}$ behaves very differently from Cr$_{\rm Ga(II)}$, and our calculations confirm this:
we find that Cr$_{\rm Ga(I)}^0$ has a $^4T_1$ ground state, consistent with the $d^7$ Tanabe-Sugano diagram for small crystal-field splitting~\cite{sugano_multiplets_1970}.

\subsection{Excited-State Properties}

To assess the optical properties of Cr$_{\rm Ga(II)}^0$, we calculate the excited states of the impurity using the constrained-occupation $\Delta$SCF approach~\cite{jones_density_1989}.
Excited states are frequently multideterminant in nature, requiring a description beyond DFT, which is a single determinant framework.
We address this issue using the approach proposed by Ziegler~\cite{ziegler_calculation_1977} and implemented within DFT by von Barth~\cite{von_barth_local-density_1979}, which entails combining the energies of several single-determinant calculations.
The equations for a $d^3$ system in an octahedral environment are given in Appendix A of Ref.~\onlinecite{shang_first-principles_2022}.
The energy of the $^2E$ excited state is given by
\begin{equation}
    \label{eq:2e_energy}
    E({^2E}) = \frac{3}{2} E({\rm mixed}) - \frac{1}{2} E({^4A_2}) \;,
\end{equation}
where $E({^4A_2})$ is the energy of the ground state, and $E({\rm mixed})$ is the energy of the broken-symmetry, single-determinant ``doublet state'' that is a mixture of the true $^2E$ excited state and the ground state [see Fig.~\ref{fig:ks}(b) for the single determinant mixed state].
We find that the $^2E$ configuration is the lowest lying excited state, with an energy 1.80~eV above the ground state.
This energy is in excellent agreement with the ZPL of 1.79~eV observed here in photoluminescence and in previous works~\cite{stehr_magneto-optical_2021,tippins_optical_1965}.
The spin-flip nature of the transition agrees with the long lifetime (on the order of ms) observed experimentally~\cite{tippins_optical_1965} and confirmed in our measurements here.

For Cr$_{\rm Ga(I)}^0$ we found a $^4T_1$ ground state and a $^2E$ excited state only 0.93~eV above the ground state.
In tetrahedral symmetry, the $^2E$ excited state is single determinant and can be trivially represented within DFT\@.
The $^2E$ excited state requires moving an electron from a $t_2$ orbital to an $e$ orbital, for which we expect coupling to phonons to be much stronger compared to the transition in Cr$_{\rm Ga(II)}^0$, leading to a much larger HR factor and a much more pronounced phonon sideband.
This effect was observed in calculations of transition metal impurities in GaN and AlN~\cite{czelej_transition-metal-related_2024}.
While Cr$_{\rm Ga(I)}$ may also be negatively charged under typical $n$-type conditions, the presence of an additional electron alters the electronic structure, making it unlikely to explain the emission.
As a result, Cr$_{\rm Ga(I)}^0$ can safely be ruled out at as the origin for the emission near 1.8~eV.

\subsection{Photoluminescence Calculations}

We calculate the photoluminescence lineshape function from first principles under the Franck-Condon approximation~\cite{lax_franckcondon_1952} using a formalism developed by Alkauskas {\it et al.}~\cite{alkauskas_first-principles_2014-1} (see Appendix~\ref{app:pl}).
Due to the high computational requirements of the HSE functional, we calculated the vibrational structure of the ground electronic state $^{4}\!A_{2}$ using the meta-GGA r$^{2}$SCAN exchange--correlation functional~\cite{furness_accurate_2020}.
This functional was selected based on its ability to accurately reproduce the frequencies of Raman-active vibrational modes in $\mathrm{Ga}_{2}\mathrm{O}_{3}$.
Supercell calculations were conducted with a $1 \times 5 \times 3$ multiple of the conventional monoclinic cell containing 300 atomic sites, using an increased 600~eV energy cutoff and Brillouin zone sampling at the $\Gamma$-point.
To compute the vibrational structure, we utilize the finite-difference method as implemented in the Phonopy software package~\cite{togo_first_2015}.

In addition to the energy for the excited state [Eq.~(\ref{eq:2e_energy})], the forces ${\bf F}_{\alpha}$ [required by Eq.~(\ref{eq:deltaQk})] are also needed to compute the lineshape function.
Taking derivatives of Eq.~(\ref{eq:2e_energy}) with respect to the atomic positions gives
\begin{equation}
    \label{eq:2e_forces}
    {\bf F}({^2E}) \approx \frac{3}{2} {\bf F}({\rm mixed}) - \frac{1}{2} {\bf F}({^4A_2}) \;.
\end{equation}
In doing so, we assume that contributions from changes in the broken-symmetry character of the mixed state are negligible~\cite{kitagawa_approximately_2007}.
The forces entering Eq.~(\ref{eq:2e_forces}) are evaluated consistently with the HSE functional.
In the equilibrium geometry of the $^4A_2$ state, ${\bf F}({^4A_2}) \approx 0$.

The computational lineshape function is shown in Fig.~\ref{fig:lineshape}.
We find overall agreement between the measured and simulated lineshape functions, allowing us to attribute most of the lower-energy peaks to the phonon sideband of the transition.
We calculate a total Huang-Rhys factor $S$ of 0.05, indicating weak electron-phonon coupling.

\section{Discussion}
\label{sec:discuss}

While there is remarkably good agreement between the {\it positions} of peaks in the calculated and experimental spectra, distinct differences in the relative {\it intensity} of various peaks occur.
A prominent example occurs in the energy range between 1.72 and 1.74~eV, where experimentally three prominent peaks occur that appear much attenuated in the simulations.
These peaks occur at energies relative to the ZPL that coincide with bulk TO phonon mode energies.
In fact, many of the peaks in the phonon sideband coincide with peaks in the bulk phonon density of states~\cite{mu_phase_2022}.
We note that these peaks also showed strong polarization dependence that differs from that of the ZPL (Fig.~\ref{fig:polarization}).

We hypothesize that the difference in intensity in the measured and simulated lineshape function may be due to
effects beyond the Franck-Condon approximation, known as the Herzberg-Teller effect~\cite{herzberg_schwingungsstruktur_1933}.
The Herzberg-Teller effect corresponds to the linear term in a series expansion of the transition dipole moment in the atomic coordinates.
In the Franck-Condon approximation, the transition matrix element, which determines the polarization, is assumed to be independent of the atomic coordinates.
As a result, the phonon sideband should exhibit the same polarization dependence as the zero-phonon line, contrary to our observations for the peaks in the range of 1.72-1.74~eV.
The Herzberg-Teller effect allows phonons to interact with the transition dipole moment, leading to differing polarization dependence as well as varied intensities with respect to the Franck-Condon approximation.

\begin{figure}
    \centering
    \includegraphics[width=\columnwidth,height=0.5\textheight,keepaspectratio]{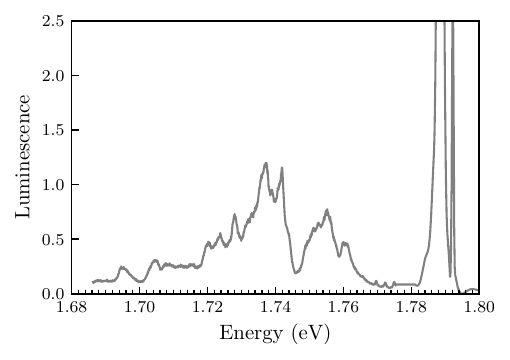}
    \caption{\label{fig:lineshape2}
        Experimental luminescence lineshape function originating from the edges of the crystal.
    }
\end{figure}

While the photoluminescence spectrum shown in Fig.~\ref{fig:lineshape} is representative of the majority of the sample, we did find that selected areas (particularly near the edges of the crystal) displayed a spectrum, shown in Fig.~\ref{fig:lineshape2}, that is distinct from the one in Fig.~\ref{fig:lineshape} and,
to our knowledge, has not been reported before.
In spite of the distinct shape, the qualitative similarity to Fig.~\ref{fig:lineshape} (including the almost identical ZPL and a phonon sideband extending over about 100 meV) suggests that this spectrum is also due to a Cr-related center.

Above, we already ruled out Cr$_{\rm Ga(I)}^0$ as the cause of the emission.
We speculate that the spectrum in Fig.~\ref{fig:lineshape2} may be due to an octahedrally coordinated Cr impurity in a secondary phase of {\gao}, given that this spectrum is only observed in some areas of the crystal.
The $\alpha$ (corundum), $\kappa$ (orthorhombic) and $\gamma$ (defective spinel) phases of {\gao} all contain octahedrally coordinated Ga sites~\cite{mu_phase_2022}.
More work is needed to verify this hypothesis.

\section{Conclusions}
\label{sec:concl}
In summary, we have reported a thorough theoretical and experimental investigation of Cr in {\bgao}.
Using first-principles calculations with a hybrid functional, we demonstrated that Cr prefers incorporating as a substitutional impurity on the octahedral Ga site in a neutral charge state.
The center has a $d^3$ electron configuration and a high-spin quartet ground state, compatible with use as a spin qubit.
It exhibits a spin-flip internal transition with a ZPL at 1.8~eV, for which we calculated the luminescence lineshape.
We performed photoluminescence experiments on samples of {\bgao} grown by the floating zone method and found good agreement between the measured and calculated lineshape functions.
Our work uncovered evidence of effects beyond the Franck-Condon approximation in the spectrum.
We also observed a previously unreported emission that we attributed to Cr in a competing phase of {\gao}.
In total, our work significantly advances the understanding of Cr in {\bgao} and opens up the potential to utilize it as a quantum defect.
Future studies isolating individual Cr impurities, including measurements of spin dynamics, photon anti-bunching, and brightness in microcavities, will guide the development and impact of {\bgao} for quantum networking, communications, and sensing.

\begin{acknowledgments}
    This work was supported via the UC Santa Barbara National Science Foundation (NSF) Quantum Foundry funded via the Q-AMASE-i program under award DMR-1906325.
    MET was supported by the U.S. Department of Energy, Office of Science, National Quantum Information Science Research Centers, Co-design Center for Quantum Advantage (C2QA) under contract number DE-SC0012704.
    SM was supported by the GAME MURI of the Air Force Office of Scientific Research (FA9550-18-1-0479).
    CVdW acknowledges a Vannevar Bush Faculty Fellowship (Office of Naval Research Award No. N00014-22-1-2808).
    SJGA acknowledges the support of the National Science Foundation Graduate Research Fellowship under Grant No. 1650114.
    LR was supported by the Research Council of Norway through the Research Project QuTe (No. 325573).
    Use was made of computational facilities purchased with funds from the NSF (CNS-1725797) and administered by the Center for Scientific Computing (CSC).
    The CSC is supported by the California NanoSystems Institute and the Materials Research Science and Engineering Center (MRSEC; NSF DMR 2308708) at UC Santa Barbara.
    This work used Bridges-2 at PSC through allocation DMR070069 from the Extreme Science and Engineering Discovery Environment (XSEDE), which was supported by National Science Foundation grant number \#1548562.
    This research also made use of the shared facilities of the NSF Materials Research Science and Engineering Center at UC Santa Barbara, Grant No. DMR-2308708.
\end{acknowledgments}

\section*{Author Contributions}
M.E. Turiansky and S. Mu contributed equally to this work.
M.E. Turiansky: Conceptualization (equal); Formal analysis (equal); Investigation (equal); Writing - original draft (equal); Writing - review \& editing (equal).
S. Mu: Conceptualization (equal); Formal analysis (equal); Investigation (equal); Writing - original draft (equal); Writing - review \& editing (equal).
L. Razinkovas: Formal analysis (equal); Investigation (equal); Writing - review \& editing (supporting).
K. Parto: Conceptualization (equal); Formal analysis (equal); Investigation (equal); Writing - review \& editing (supporting).
S.D. Patel: Formal analysis (equal); Investigation (equal); Writing - review \& editing (supporting).
S. Doan: Formal analysis (equal); Investigation (equal); Writing - review \& editing (supporting).
G. Pokharel: Conceptualization (equal); Formal analysis (equal); Investigation (equal); Writing - review \& editing (supporting).
S.J. Gomez Alvarado: Formal analysis (equal); Investigation (equal); Writing - review \& editing (supporting).
S.D. Wilson: Conceptualization (equal); Formal analysis (equal); Writing - review \& editing (supporting); Supervision (equal); Funding acquisition (equal).
G. Moody: Conceptualization (equal); Formal analysis (equal); Writing - review \& editing (supporting); Supervision (equal); Funding acquisition (equal).
C.G. Van de Walle: Conceptualization (equal); Formal analysis (equal); Writing - review \& editing (equal); Supervision (equal); Funding acquisition (equal).

\section*{Data Availability}
The computational data that supports the findings of this study are archived in the NOMAD repository~\cite{draxl_nomad_2019} at \url{[TO BE INSERTED]}.
The experimental data that supports the findings of this study are available from the corresponding author upon reasonable request.

\appendix

\section{Theory of the Luminescence Lineshape}
\label{app:pl}
We calculate the luminescence lineshape for the ${^{2}\!E} \rightarrow {^{4}A_{2}}$ transition using the methodology outlined in Refs.~\onlinecite{alkauskas_first-principles_2014-1,razinkovas_vibrational_2021}.
Employing the Franck--Condon approximation and assuming the zero-temperature limit (0~K), the normalized luminescence intensity $L(\hbar\omega)$ is expressed as:
\begin{eqnarray}
  L(\hbar\omega) = C \omega^{3} A(\hbar\omega),
\end{eqnarray}
where $C$ is a normalization constant which ensures that $\int L(\hbar\omega) \mathrm{d}\hbar\omega = 1$.
The function $A(\hbar\omega)$ represents the spectral function which characterizes the phonon sideband:
\begin{align}
  A(\hbar\omega) =
  &
    \sum_n
    \left|
    \braket{\chi^{g}_{n} \vert \chi^{e}_{0}}
    \right|^2
    \delta
    \left(
    E_{\mathrm{ZPL}} + \varepsilon^{g}_{n} - \varepsilon^{e}_{0} -\hbar\omega
    \right).
\end{align}
In the equation above, $\ket{\chi^{i}_{m}}$ represents the harmonic vibrational wavefunction of electronic state $i$, characterized by vibrational quantum number $m$ and the associated vibrational energy $\varepsilon^{i}_{m}$.
The estimation of $A(\hbar\omega)$ requires computing multidimensional overlap integrals between vibrational wavefunctions $\braket{\chi^{g}_{n} \vert \chi^{e}_{0}}$.
This process is simplified by employing the equal-mode approximation~\cite{markham_interaction_1959,razinkovas_vibrational_2021}:
we assume that the shapes and frequencies of vibrational modes of the initial and final electronic state are identical, requiring the calculation of only the ground-state vibrational structure and linear change in the adiabatic potential energy surfaces.
In this formulation, the strength of electron--phonon interaction is described by the spectral density of electron--phonon coupling:
\begin{equation}
  \label{eq:S}
  S(\hbar\omega) = \sum_{k} S_{k} \delta(\hbar\omega_{k} - \hbar\omega).
\end{equation}
Here, $S_k = {\omega_k \Delta Q_k^2}/{2\hbar}$ is the partial Huang--Rhys factor, quantifying the coupling strength for vibrational mode~$k$~\cite{huang_kun_theory_1950}.
In evaluating Eq.~\eqref{eq:S}, Dirac delta functions $\delta$ are replaced with Gaussian functions of 1.2~meV widths to simulate broadening effects.
$\Delta Q_k$ describes the alteration of equilibrium geometry upon optical transition along the direction of vibrational mode $\boldsymbol{\eta}_{k}$ (in mass-weighted form) of frequency $\omega_k$.
We calculate $\Delta Q_k$ using the force $\boldsymbol{F}_{\alpha}$ exerted on each atom $\alpha$ of mass $M_\alpha$, induced by the electronic transition~\cite{razinkovas_vibrational_2021}:
\begin{equation}
  \Delta Q_k = \frac{1}{\omega_k^2} \sum_{\alpha} \frac{\mathbf{F}_{\alpha}}{\sqrt{M_{\alpha}}} \cdot \boldsymbol{\eta}_{k;\alpha}.
  \label{eq:deltaQk}
\end{equation}
With spectral density in hand, the spectral function $A(\hbar\omega)$ is then determined using the generating function approach~\cite{lax_franckcondon_1952,alkauskas_first-principles_2014-1,razinkovas_vibrational_2021}:
\begin{eqnarray}
  && A(\hbar \omega) = \frac{1}{2\pi} \int_{-\infty}^{\infty} e^{i\omega t} G(t)
     e^{-\gamma |t|} \, \mathrm{d}t,
     \label{eq:A2}
  \\
  && G(t) = \exp\left[-\frac{iE_{\mathrm{ZPL}}t}{\hbar} - S + \int e^{i\omega
     t} S(\hbar \omega)\,\mathrm{d}\hbar\omega\right].
     \label{eq:G}
\end{eqnarray}
In this formulation, $G(t)$ serves as the generating function for luminescence.
The parameter $\gamma$ accounts for homogeneous broadening effects that are not captured by the theory and is adjusted to match the experimentally observed ZPL width.
We utilize $\gamma = 0.007$~meV to ensure alignment between the theoretical and experimental zero-phonon linewidths.
$S=\sum_{k}S_{k}$ is the total Huang--Rhys factor, which quantifies the average number of excited phonons during the optical transition.

\bibliographystyle{apsrev4-2}

\end{document}